\documentclass{osa-article}

\journal{oe}

\usepackage{hyperref}
\hypersetup{
    colorlinks=true,
    linkcolor=blue,
    filecolor=blue,      
    urlcolor=blue,
}
\newcommand{\bi}{\begin{itemize}}
\newcommand{\ei}{\end{itemize}}
\newcommand{\ben}{\begin{enumerate}}
\newcommand{\een}{\end{enumerate}}
\newcommand{\be}{\begin{equation}}
\newcommand{\ee}{\end{equation}}
\newcommand{\bea}{\begin{eqnarray}} 
\newcommand{\eea}{\end{eqnarray}}
\newcommand{\ba}{\begin{align}} 
\newcommand{\ea}{\end{align}}
\newcommand{\bse}{\begin{subequations}} 
\newcommand{\ese}{\end{subequations}}
\newcommand{\bc}{\begin{center}}
\newcommand{\ec}{\end{center}}
\newcommand{\bfi}{\begin{figure}}
\newcommand{\efi}{\end{figure}}

\newcommand{\bmp}[1]{\begin{minipage}{#1}}
\newcommand{\emp}{\end{minipage}}

\newcommand*\dif{\mathop{}\!\mathrm{d}}

\usepackage{xcolor}

\renewcommand{\eqref}[1]{Eq.~(\ref{eq:#1})}

\articletype{Research Article}

\begin{document}
\title{Maximal nighttime electrical power generation via optimal radiative cooling}
\author{Lingling Fan$^{1, 3}$, Wei Li$^1$, Weiliang Jin$^1$, Meir Orenstein$^2$, Shanhui Fan$^{1, *}$}
 \address{$^1$Department of Electrical Engineering, Ginzton Laboratory, Stanford University, Stanford, California 94305, USA\\
$^2$Department of Electrical Engineering, Technion-Israel Institute of Technology, 32000 Haifa, Israel
 }
\email{
$^{3}$llfan@stanford.edu\\
$^*$shanhui@stanford.edu}


\begin{abstract}
We present a systematic optimization of nighttime thermoelectric power generation system utilizing radiative cooling. We show that an electrical power density $>2$ W/m$^2$, two orders of magnitude higher than the previously reported experimental result, is achievable using existing technologies. This system combines radiative cooling and thermoelectric power generation and operates at night when solar energy harvesting is unavailable. The thermoelectric power generator (TEG) itself covers less than 1 percent of the system footprint area when achieving this optimal power generation, showing economic feasibility. We study the influence of emissivity spectra, thermal convection, thermoelectric figure of merit and the area ratio between  the TEG and the radiative cooler on the power generation performance. We optimize the thermal radiation emitter attached to the cold side and propose practical material implementation. The importance of the optimal emitter is elucidated by the gain of 153$\%$ in power density compared to regular blackbody emitters.
\end{abstract}

\section{Introduction}

\par The rapid growth of world population and industrial development poses a major threat to the global energy supply, causing escalatory
environmental degradation and social unrest \cite{doi:10.1146/annurev.energy.30.050504.144228, Chu2017, Shindell2019}. To address the challenge, solar energy harvesting techniques, such as photovoltaics, thermal photovoltaics, and solar thermal techniques\cite{Schwede2010, Lenert2014, Cooper2018, Omair15356}, are sustainable alternatives. Even so, the electrical power demand for lighting peaks during nighttime. {\cite{ZHOU2018499}. Moreover, due to the uneven distribution of sunlight, the access to solar energy can be rather limited for months in many places on the Earth.} A feasible approach to mitigate this discrepancy is to develop a passive system that can generate work during nighttime. An off-grid power generator needs to be of low-cost, ruling out many traditional approaches. In addition to lighting, a modular energy source can benefit a large variety of off-grid sensors (agriculture, environmental, security), digital communications and many other applications. 

    \par Realization of passive nighttime power generation at a level of 1 W/m$^2$ is facing a few challenges. First, efficient harvesting of environmental energy to the generator input; second, efficient utilization of the harvested energy to generate power; third, efficient dumping of excessive heat from the generator output. The available energy source at nighttime is the heat contents of the atmosphere that can be collected by free air convection into the generator. {With typical temperature difference between the ambient and hot side of few degrees K and the free air convection coefficient of 8$\sim$10 W/m$^2$K \cite{RAMAN20192679}, the input power density provided to the generator is limited to on the order of 10 W/m$^2$ }- which, with the typical power conversion efficiency of TEG, cannot provide the desired 1 W/m$^2$  power density, thus requiring significant enhancement. Heat dumping is another challenge. Luckily we have a {stably low-temperature cold} sink, the outer space, which is maintained at about 3 K and is everywhere available and evenly distributed. Previous works utilizing the cold outer space has enabled passive radiative cooling of the surface to well below the ambient air temperature \cite{doi:10.1063_1.329270, doi:10.1021/nl903271d, doi:10.1021/nl4004283, Raman2014, Chen2016, Zhai1062, Mandal315}. With such temperature gradient, it is possible to set up a heat engine to extract work between the ambient environment and outer space \cite{Byrnes3927, BuddhirajuE3609}. A proof-of-concept experiment \cite{RAMAN20192679} developed a device that couples the cold side of a TEG to a sky-facing blackbody surface that radiates heat to the cold of space and has its hot side heated by the surrounding air, enabling electrical power generation of 25 mW/m$^2$ at night. Although this demonstration of nighttime electrical power generation is remarkable, it is not sufficient to fulfill energy demand of many applications mentioned above. The power generation performance in \cite{RAMAN20192679} can be possibly improved by tackling the following two challenges: first, the blackbody emitter of the radiative cooler in \cite{RAMAN20192679} does not provide the optimal cooling throughput, since it absorbs heat power at frequencies/angles where the atmospheric emission is dominant; second, its thermal structure is adverse to forming thermoelectric current due to the presence of excessive parasitic heat loss. On the other hand, previous works on daytime high-performance radiative cooling have explored the optimization of spectro-angular selective emitters \cite{Raman2014, doi:10.1021/acsphotonics.7b01136, Bhatia2018,Zhou2019} and heat transfer condition \cite{Chen2016, CHEN2019101}. However, to the best of our knowledge, {engineering the two conditions} has not been investigated to optimize the nighttime electrical power generation performance.
\par In this work, we show that, with a systematic optimization, the nighttime thermoelectric power generation of 2.2 W/m$^2$ is achievable using current technologies. This power density presents an improvement of 88 times compared to the power density demonstrated in \cite{RAMAN20192679}. {The optimal performance is enabled by comprehensive optimization of radiative coolers (spectro-angular selective emissivity and the area ratio over the TEG), environmental convection (hot and cold side convection coefficient) and thermoelectric figure of merit ZT factor.} {Throughout the paper, we fix the single thermocouple effective area $A_{\rm TC}$ and change the TEG area $(A_{\rm TE} = NA_{\rm TC})$ by varying the number of thermocouples $N$.} We demonstrate that the optimized radiative cooler with spectro-angular selectivity can yield high-performance power generation in various environmental and thermoelectric device conditions. We show that, when employing a Carnot engine under the same conditions, the upper limit of nighttime power density generated at ambient temperature 300 K is about 6.4 W/m$^2$. This indicates that our optimal TEG design is close to the thermodynamic limit (the limit of usable electrical power generation from TEG is half of that of the Carnot engine).

\par The paper is organized as follows: In Section \ref{Sec_2}, we present the theory and design principles of a
thermoelectric power generator at nighttime coupled with optimal radiative cooler. We give a concrete design with typical experimental parameters and verify that the optimal emitter outperforms the conventional blackbody emitter in Section \ref{sec_3}. In Section \ref{secexper}, we present a feasible thermoelectric power generation system design that generates 2.2 W/m$^2$ with a vacuum enclosure around the cold emitter, using a commercially available TEG. In Section \ref{sec4}, we analyze the influence of environmental condition, thermoelectric figure of merit and radiative area ratio on the power generation performance. We discuss the ultimate limit of power generation when a Carnot engine operates between the ambient and the radiative cooler to extract work in Section \ref{sec5} and conclude in Section \ref{sec_6}.

\begin{figure}[ht]
\centering\includegraphics[scale = 0.3]{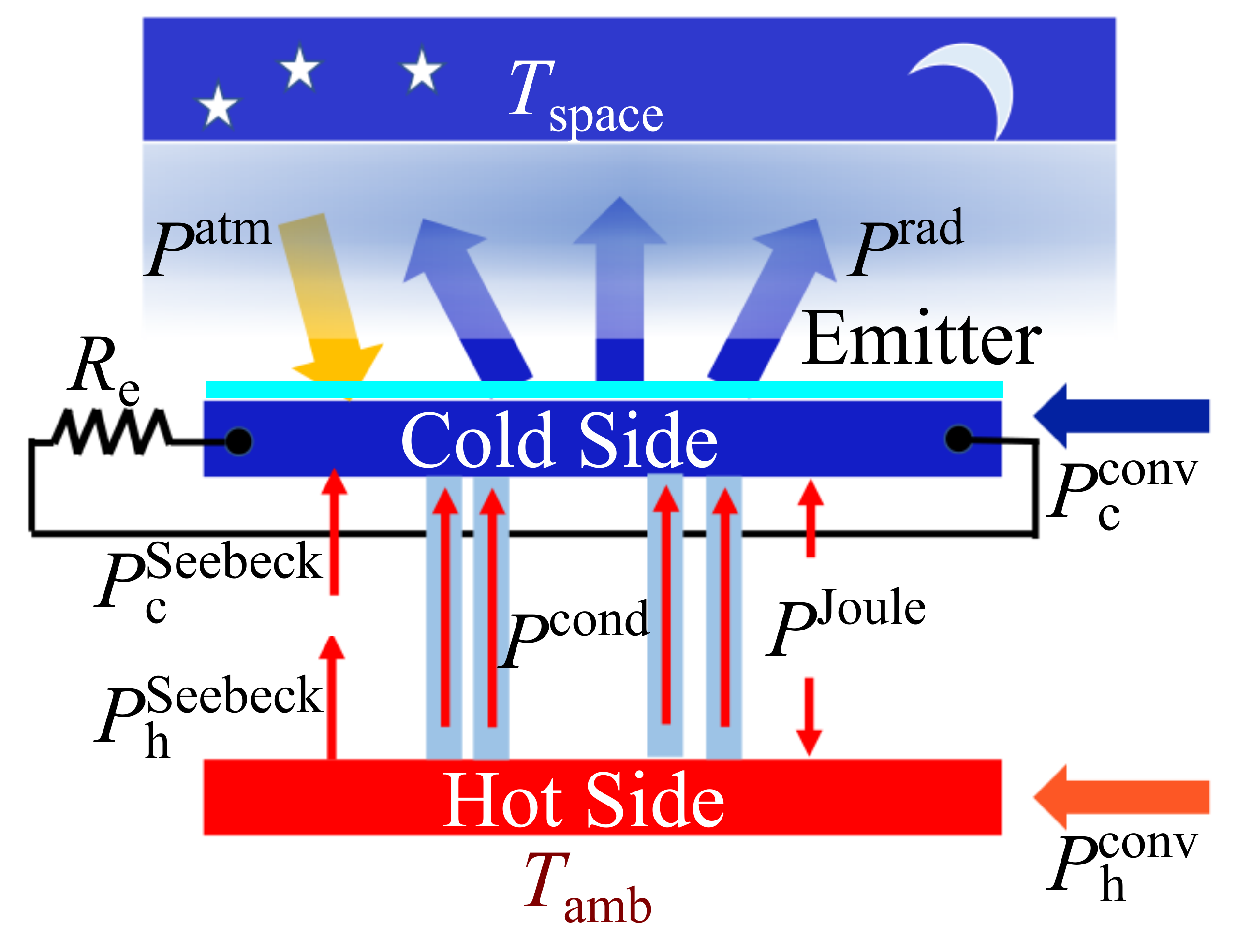}
\caption{Schematic setup of the TEG utilizing radiative cooling.}
\label{fig_1_first}
\end{figure}
\section{Analysis of nighttime thermoelectric power generation}\label{Sec_2}
  \par We start by analyzing the thermodynamic model of nighttime thermoelectric power generator that is schematically shown in Fig. \ref{fig_1_first}. We integrate the cold side of TEG at temperature $T_{\rm c}$ with a radiative cooler, whose radiative properties are described by the emissivity $\epsilon(\lambda,\theta)$ at wavelength $\lambda$ and incident angle from the normal direction $\theta$. The cold side is exposed to the clear night sky and subject to atmospheric irradiance that depends on the ambient temperature $T_{\rm amb}$. The temperatures of the cold and hot sides of the TEG, $T_{\rm c}$ and $T_{\rm h}$, can be obtained by solving the steady-state power balance of the two sides:
  \begin{align}
    P^{\rm rad} -  P^{\rm atm}- P^{\rm cond} - P^{\rm conv}_{\rm c} - P^{\rm Joule}-P_{\rm c}^{\rm Seebeck}&=0,
    \label{cold_side_balance}\\  P^{\rm cond}-P_{\rm h}^{\rm conv}-P^{\rm Joule}+P_{\rm h}^{\rm Seebeck}&= 0,
    \label{net_hot}
\end{align}
\noindent{where $P^{\rm rad}= A_{\rm c}\int\dif\Omega\cos\theta\int_0^{\infty}\dif{\lambda}I_{\rm BB}(T_{\rm c}, \lambda)\epsilon(\lambda, \theta)
$ is the power radiated out by the radiative cooler,} $P^{\rm atm} =A_{\rm c} \int\dif\Omega\cos\theta\int_0^{\infty}\dif{\lambda}I_{\rm BB}(T_{\rm amb}, \lambda)\epsilon(\lambda, \theta)\epsilon_{\rm atm}(\lambda, \theta)$ is the power absorbed in the cold side stemming from the atmosphere radiation. Here, $A_{\rm c}$ is the cold-side surface area of the structure. $\int\dif\Omega = 2\pi\int_0^{\pi/2}\dif{\theta}\sin\theta$ is the angular integral over a hemisphere. $I_{\rm BB}(T, \lambda) = \frac{2hc^2}{\lambda^5}\frac{1}{\exp[{hc/(\lambda k_{\rm B}T)}] - 1}$ is the spectral radiance of a blackbody at temperature $T$, where $k_{\rm B}$ is the Boltzmann constant, $c$ is the speed of light and $h$ is the {{Planck constant}}. The angle-dependent spectral emissivity of the atmosphere is given by \cite{doi:10.1063/1.329270}: $\epsilon_{\rm atm}(\lambda, \theta) = 1- t(\lambda)^{1/\cos\theta}$, where $t(\lambda)$ is the atmospheric transmittance in the zenith direction \cite{nasa_memo,obserb_web}. $P^{\rm cond} ={ (T_{\rm h} - T_{\rm c})}/{R_{\rm TE}}$ represents the internal parasitic heat transfer from the hot to the cold side due to conduction with $R_{\rm TE}$ being the thermal resistance of the TEG structure. $P^{\rm conv}_{\rm c} = A_{\rm c}h_{\rm c}(T_{\rm amb} - T_{\rm c})$ and $P_{\rm h}^{\rm conv} =A_{\rm h} h_{\rm h}(T_{\rm amb} - T_{\rm h})$ are the heat transfer from the ambient to the cold and hot side respectively due to air convection, where $A_{\rm h}$ is the hot-side surface area of the structure, and $h_{\rm h}$ and $h_{\rm c}$ are the air convective
heat transfer coefficients due to the
contact of air adjacent to the hot side of TEG and the radiative cooler, respectively.  
\par Additionally, there is heat power related to the Joule heating and Seebeck effect in Eqs. (\ref{cold_side_balance}) and (\ref{net_hot}). $P^{\rm Joule}=\frac{N}{2}I^2R_{\rm np}$ is the heat provided to either the hot and cold side due to Joule heating of the internal resistance. Here following standard treatment \cite{macro_book1, theromobok_1} for simplicity we assume that such heat is provided equally to the hot and the cold sides. The current is $I = NS_{\rm np}(T_{\rm h} - T_{\rm c})/(NR_{\rm np} + R_{\rm e})$, where $N$ is the number of thermocouples in the TEG and $S_{\rm np}$ is the Seebeck coefficient of a single junction. $R_{\rm e}$ and $R_{\rm np}$ are the external and single thermocouple's internal electrical resistances, respectively. $P_{\rm h}^{\rm Seebeck}= NS_{\rm np}T_{\rm h}I$ and $P_{\rm c}^{\rm Seebeck}= NS_{\rm np}T_{\rm c}I$ are the heat outflow and inflow of the hot and cold side due to the Seebeck effect, respectively. The hot side is assumed to have a very low emissivity, thus the radiated power of the hot side is negligible. This assumption is widely applied in radiative cooling devices \cite{Zhu:14, Zhu12282, RAMAN20192679} and is applicable here as well. In our realization, the hot side is immersed in air and separated from other thermal reservoirs such as the ground or roof, so the heat convection from air is the sole power source of the system. {{We neglect other parasitic heat gains or losses, such as conduction from the support, due to their small quantity compared with the dominating heat flows.}} Given these assumptions, Equation (\ref{net_hot}) gives $\frac{(T_{\rm h} - T_{\rm c})}{R_{\rm TE}}+\frac{Z(T_{\rm h} - T_{\rm c})}{2R_{\rm TE}}T_{\rm h}-\frac{Z(T_{\rm h} - T_{\rm c})^2}{8R_{\rm TE}}-A_{\rm h} h_{\rm h}(T_{\rm amb} - T_{\rm h})=0$, where $Z = \frac{NS_{\rm np}^2R_{\rm TE}}{R_{\rm np}}$. Combined with Eq. (\ref{cold_side_balance}), both $T_{\rm c}$ and $T_{\rm h}$ can be solved.
\par
\begin{figure}[ht]
\centering\includegraphics[scale = 1.5]{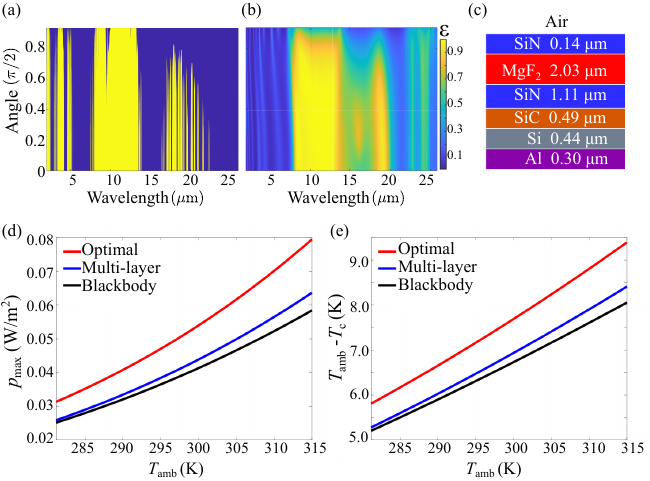}
\caption{Nighttime power generator with selective thermal emitter for improved thermoelectric power generation. (a) The ideal emissivity (Eq. (\ref{optim_emiss})) for optimal thermoelectric power generation at ambient temperature of 300 K, cooling down the emitter to $T_{\rm c} = 292.3$ K. (b) The emissivity of the optimized multi-layer at ambient temperature of 300 K, cooling down the emitter to $T_{\rm c} = 293.1$ K. (c) The material composition and thicknesses for the multi-layer structure with spectro-angular selectivity depicted in (b). (d) Output power density $p_{\rm max}$ of the above three emitters at different ambient temperatures. (e) Temperature difference between the radiative cooler and ambient for the three emitters. The other parameters of the TEG system are assumed to be the same as \cite{RAMAN20192679}. }
\label{fig_1}
\end{figure}

 To maximize electric power generation for a given temperature difference $T_{\rm h} - T_{\rm c}$, we apply the load-matching condition where $R_{\rm e} = NR_{\rm np}$. The maximum power density $p_{\rm max}$ is obtained directly from Eqs. (\ref{cold_side_balance}) and (\ref{net_hot}) as the difference between the net external power delivered to the hot side and leaving the cold side: 

\begin{align}
p_{\rm max}=\frac{NS_{\rm np}^2(T_{\rm h} - T_{\rm c})^2}{4R_{\rm np} A_{\rm c}} .\label{Power_max}
\end{align} In the proof-of-concept experiment with blackbody emitter reported in \cite{RAMAN20192679}, the power generation system parameters were: $T_{\rm amb}\sim281$ K, $h_{\rm h}=10$ W/(m$^2$K), $h_{\rm c}\sim7$ W/(m$^2$K), $R_{\rm TE} = 2.5$ K/W, $A_{\rm c} = A_{\rm h} = 0.01\pi$ m$^2$, $N = 127$, $S_{\rm np}= 210.769$ $\mu$V/K, $R_{\rm np}$ = 0.007 $\Omega$. The single thermocouple effective area is $A_{\rm TC} = A_{\rm TE}/N$, where the TEG area $A_{\rm TE}$ is $30\times 30$ mm$^2$ from Marlow TG12-4-01LS used in \cite{RAMAN20192679}. The temperature difference obtained in this experiment was $T_{\rm h} - T_{\rm c} =1.98$ K and the generated electrical power density $p_{\rm max}=0.025$ W/m$^2$. Our model reproduced faithfully the experimental result, as shown in Fig. \ref{fig_1}(d, e). As we will demonstrate below, this power generation performance can be improved significantly by optimally engineering the radiative cooler emissivity spectra, the area ratio between the TEG and the radiative cooler, the thermoelectric figure of merit, and environmental convection conditions.

\section{Optimization of the radiative cooler emissivity}\label{sec_3}
    \par To maximize the power density generation $p_{\rm max}$ of the TEG, the temperature of its cold side should be decreased as much as possible as evidenced by Eq. (\ref{Power_max}), which makes the optimal radiative cooler a critical element in the system. The power density balance of the cooler can be written as: \begin{align}
        \Delta p_{\rm r}(T_{\rm c}) =  \Delta p_{\rm par}(T_{\rm c}),\label{lhs_rhs_eq} \end{align}where $\Delta p_{\rm r}(T_{\rm c})=\int\dif\Omega\cos\theta\int_0^{\infty}\dif{\lambda}[I_{\rm BB}(T_{\rm c}, \lambda)- I_{\rm BB} (T_{\rm amb}, \lambda)\epsilon_{\rm atm}(\lambda, \theta)]\epsilon(\lambda, \theta)
  $ is the net radiative power density from the cold side and $
         \Delta p_{\rm par}=h_{\rm c}(T_{\rm amb} - T_{\rm c})+\frac{1}{ A_{\rm c}}\left[\frac{T_{\rm h} - T_{\rm c}}{R_{\rm TE}}+\frac{Z(T_{\rm h} - T_{\rm c})^2}{8R_{\rm TE}}+\frac{Z(T_{\rm h} - T_{\rm c})}{2R_{\rm TE}}T_{\rm c}\right]
$ is the parasitic heat transfer density into the cold side. The $T_{\rm c}$ solution is uniquely determined by the intersection of the two monotonous functions $ \Delta p_{\rm r}(T_{\rm c})$ and $\Delta p_{\rm par}(T_{\rm c})$. This derivation highlights two important factors that influence the performance of the power generation: the control of the cold side emissivity $\epsilon(\lambda, \theta)$, as well as various parameters related to the TEG setup which represents the effective heat transfer coefficient of all parasitic heat transfer and controls the cold side temperature. 
\par In this section, we focus on the emissivity design. For a given $T_{\rm c}$, the optimal emissivity spectrum $\epsilon(\lambda, \theta)$ should maximize the cooling power by filtering out the negative contribution of the integral in $ \Delta p_{\rm r}$. This is achieved by assigning $\epsilon(\lambda, \theta) = 1$ when the amount of power radiated out from the cooler is larger than the power it absorbs from the atmosphere radiation and otherwise $\epsilon(\lambda, \theta) = 0$. The optimal emissivity should conform with:
\begin{align}
    \epsilon(\lambda, \theta) = \Theta[I_{\rm BB}(T_{\rm c},\lambda) - \epsilon_{\rm atm}(\lambda, \theta)I_{\rm BB}( T_{\rm amb},\lambda)], \label{optim_emiss}
\end{align}
where $\Theta$ is the unit step function, and $T_{\rm c}$ is solved self-consistently from Eqs. (\ref{cold_side_balance}), (\ref{net_hot}) and (\ref{optim_emiss}). The spectral selectivity of the optimal cooler gives strong emission at frequencies where the atmospheric absorption (in the range of 8-13 $\mu$m) and the ozone layer reflection ($\sim9.5$ $\mu$m) are relatively smaller. The angular selectivity of the emitter also prevents emissions at large incident angular ranges where the sky is mostly opaque and the downward sky radiation is intensive. The optimal emissivity spectrum $\epsilon(\lambda, \theta)$ is shown in Fig. \ref{fig_1}(a) at nighttime ambient temperature 300 K and other conditions the same as \cite{RAMAN20192679}. This optimal emissivity cools down the cold side to $T_{\rm c} = 292.3$ K as determined by Eqs. (\ref{cold_side_balance}), (\ref{net_hot}) and (\ref{optim_emiss}) and results in a power generation of 0.054 W/m$^2$, higher compared to 0.041 W/m$^2$ for a blackbody emitter under the same conditions.
\par To implement an approximated optimized structure of radiative cooler for electrical power generation, we consider the use of a multi-layer emitter. Similar multi-layer structures have been applied in radiative cooling and thermal management designs in recent years \cite{Raman2014, Chen2016, doi:10.1021/acsphotonics.7b00089, Li2018}. To find the multi-layer structure for high-performance electrical power generation, we perform a broadband optimization \cite{doi:10.1021/acsphotonics.7b01136, nlopt} with the radiative power density $ \Delta p_{\rm r}$ as the merit function with $T_{\rm c} = 292.3$ K which is the temperature of the aforementioned optimal emitter. To achieve maximal power generation with a five-layer structure, we employ a diverse set of materials so that the multi-layer structure can have an emissivity that approximately matches the optimal emissivity spectrum in Fig. \ref{fig_1}(a), taking into account fabrication feasibility. The superstrate is chosen as air, and the structure is placed on top of 300 nm of Al attached to the cold side of the TEG. We select from the following ten common
dielectric materials: Al$_2$O$_3$, HfO$_2$, MgF$_2$, SiC, SiN, SiO$_2$, TiO$_2$, Ta$_2$O$_5$, Si, Si$_3$N$_4$. The thick Al is {{opaque}} to thermal radiation, thus the emissivity spectrum is $\epsilon(\lambda, \theta) = 1- R(\lambda, \theta)$, where $R(\lambda, \theta) = \frac{1}{2}(R_{\rm s}(\lambda, \theta) + R_{\rm p}(\lambda, \theta) ) $ is the average
reflectivity of the $s-$ and $p-$polarized light, both of which were calculated by applying the analytical surface impedance method on a multi-layer structure as
described in \cite{haus_book}. Using the emissivity, we calculate the net radiative cooling power density $ \Delta p_{\rm r}$ to evaluate the electrical power generation for the multi-layer device. In Fig. \ref{fig_1}(b), we plot the emissivity spectrum of the calculated multi-layer structure, which demonstrates the selectivity on wavelength and incident angle that is favorable for thermoelectric power generation. The emissivity spectrum of the designed emitter approximates that of the theoretically optimal spectrum specified in Fig. \ref{fig_1}(a). The material composition and thickness constituents of the optimized emitter are shown in Fig. \ref{fig_1}(c). 

\par To assess the power generation performance of this device, in Fig. \ref{fig_1}(d), we compare the electrical power density of the optimal emitter described by Eq. (\ref{optim_emiss}), the optimized multi-layer emitter shown in Fig. \ref{fig_1}(c), and the blackbody device of Ref. \cite{RAMAN20192679}, under the same environmental and TEG conditions for a range of ambient temperatures. {This demonstrates the possible performance variations in different locations on the Earth.} We find that, at all
temperatures, our optimized multi-layer structure generates higher electrical power than the blackbody emitter. Nevertheless, the optimized multi-layer structure is outperformed by the optimal emitter shown in Eq. (\ref{optim_emiss}) which leaves space for even better emitter designs. Next, we explore the working principle of the optimal emitter, by presenting in Fig. \ref{fig_1}(e) the temperature reduction of the cooler from the ambient. This verifies that, at all ambient temperatures, the optimal emitter lowers the cooler temperature further than the multi-layer and blackbody emitter. In the future, we envision that by exploring adjoint optimization method \cite{Jin:17, Jin:18, PhysRevB.99.041403} with diverse geometrical and material combinations, one may potentially achieve emissivity design closer to the ideal goal of Eq. (\ref{optim_emiss}), and consequently improve the power generation performance.

\begin{figure}[ht]
\centering\includegraphics[scale = 1.5]{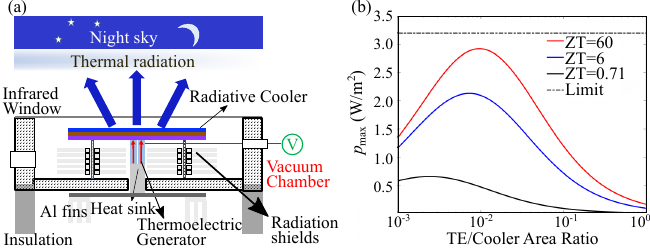}
\caption{The proposed system for optimal power generation at nighttime. (a) Schematic of the setup. (b) The output power density $p_{\rm max}$ as a function of thermoelectric to radiative cooler area ratio for various thermoelectric figure-of-merit values, as well as the limit determined by half of the Carnot engine extracted power density, with $h_{\rm c} = 10^{-3}$ W/(m$^2$K) and $h_{\rm h} = 10^2$ W/(m$^2$K) at the ambient temperature of 300 K. }
\label{fig_4}
\end{figure}
\section{Feasible optimal design of high-performance nighttime thermoelectric power generation }\label{secexper}

    \par We present in Fig. \ref{fig_4} a feasible nighttime thermoelectric generator design that optimally generates 2.2 W/m$^2$ power density. Our optimization is enabled by {{a state-of-the-art TEG (ZT = 6) that is suitable for the temperature range of nighttime implementation}} \cite{TEG_record} and we compare it with more futuristic ZT = 60 TEG \cite{Shimizu2019_1} and experimentally implemented ZT = 0.71 TEG \cite{RAMAN20192679}. Here, ZT $=N S_{\rm np}^2R_{\rm TE}T_{\rm amb}/R_{\rm np}$ is the thermoelectric figure of merit at ambient temperature of 300 K. The system is enclosed by engineered thermal convection conditions: the vacuum cold-side environment can achieve negligible parasitic heating by air convection with $h_{\rm c}$ = $10^{-3}$ W/(m$^2$K), and the heat sink \cite{article_maranzana, Hireholi2013ExperimentalAT} attached to the hot side of the TEG can increase the effective area for convection by a factor of 10 for still air convection, to yield effectively $h_{\rm h}$ = $10^{2}$ W/(m$^2$K). Furthermore, radiation shields and isolation pegs are used to reduce the radiation and conduction loss
through the backside of the emitter. 
\par To optimize the number of thermocouples in a rooftop application setting, where the system footprint area is assumed to be $A_{\rm c}=A_{\rm h} = 1$ m$^2$, we study the thermoelectric power density $p_{\rm max}$ as a function of the area ratio of the TEG to the radiative cooler $A_{\rm TE}/A_{\rm c}$ in Fig. \ref{fig_4}(b). The power density is calculated as the generated power divided by this footprint area, and the convection coefficients of the hot and cold sides are chosen to be the same as in Fig. \ref{fig_4}(a). Each of the emitters are designed to be optimal according to Eq. (\ref{optim_emiss}). For all of the three emitters, the output power density $p_{\rm max}$ peaks at a respective $A_{\rm TE}/A_{\rm c}$ area ratio. For the experimental ZT = 0.71 case \cite{RAMAN20192679}, the optimal output power density $p_{\rm max}$ is  0.67 W/m$^2$ when the TEG consists of 339 thermocouples and corresponding area ratio $A_{\rm TE}/A_{\rm c}$ is 0.0024. For the available ZT = 6 case \cite{TEG_record}, the optimal output power density $p_{\rm max}$ is  2.2 W/m$^2$ obtained with 1001 thermocouples and corresponding area ratio $A_{\rm TE}/A_{\rm c}$ is 0.007. For the futuristic ZT = 60 case \cite{Shimizu2019_1}, the optimal output power density $p_{\rm max}$ is  2.92 W/m$^2$ with 1351 thermocouples and corresponding area ratio $A_{\rm TE}/A_{\rm c}$ of 0.0096. For all three cases the maximum power density is achieved with the TEG area less than 1 percent of the radiative cooler area. This is an important observation since the TEG is the most expensive part of the system. The upper bound of the thermoelectric power density generation on the load resistor is half of that obtained by an ideal Carnot engine (see Section \ref{sec5}). Our results for a technically achievable design are not far from this Carnot limit of 3.2 W/m$^2$,  denoted by a dashed line in Fig. \ref{fig_4}(b). We also note that producing 2.2 W/m$^2$ at night from an environmental source outperforms thermal energy harvesting from human body, as well as energy from radio frequency, and is comparable with other small-scale ambient energy harvesting techniques such as wind \cite{ti_report, doi:10.1177/1687814017696210, sensor_report}. {The current configuration in Fig. \ref{fig_4} may operate also at daytime in the reverse direction, where the Sun is heating the emitter with strong visible emissivity. We are examining the optimal designs for operation in both day and nighttime in an upcoming publication \cite{lf_new_unknown}.}
\begin{figure}[ht]
\centering\includegraphics[scale = 1.5]{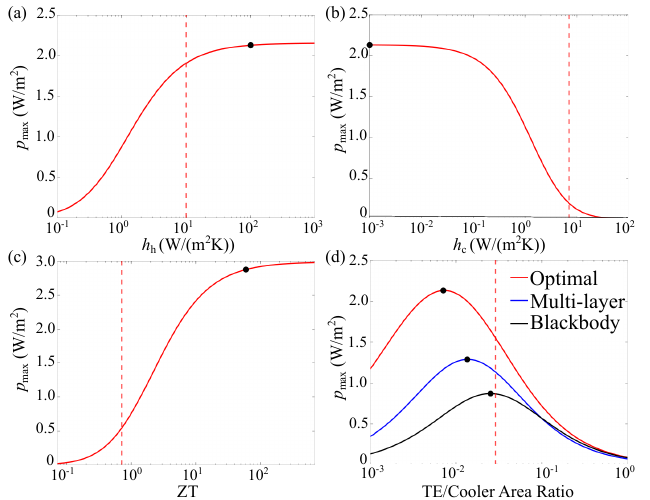}
\caption{The output power density $p_{\rm max}$ as a function of system parameters: convection coefficients (panels a and b), thermoelectric figure of merit (panel c) and TEG/radiative cooler area ratio (panel d) at the ambient temperature of 300 K. In each panel, we study the impact of a respective parameter at a fixed value of the other parameters that are optimized at ZT = 6 (section \ref{secexper}). The red dashed line in each panel indicates the value of the respective parameter in the experiment \cite{RAMAN20192679}: $h_{\rm h}=10$ W/m$^2$/K, $h_{\rm c}$ = 7 W/m$^2$/K, ZT $ = 0.71$, and $A_{\rm TE}/A_{\rm c}= 0.0286$. The black point in each panel denotes the parameter used for the optimal power generation performance. }
\label{fig_3}
\end{figure}
\section{Parameter influence on the system performance}\label{sec4}

\par In addition to the critical role of the optimal emitter, as shown in Fig. \ref{fig_3}, we perform a detailed study of the system parameters including convection coefficients, thermoelectric figure of merit and TEG/radiative cooler area ratio. We consider the TEG with ZT = 6 of Fig. \ref{fig_4}(b) at ambient temperature $T_{\rm amb} = 300$ K integrated with radiative cooler with the optimized emitter as a baseline, and we scan values of each single parameter - leaving the other parameters unchanged. A red dashed line marks the value of each specific parameter used in the experiment of \cite{RAMAN20192679}: $h_{\rm h}=10$ W/m$^2$/K, $h_{\rm c}$ = 7 W/m$^2$/K, ZT $ = 0.71$, and $A_{\rm TE}/A_{\rm c}= 0.0286$.

\par As can be seen in Fig. \ref{fig_3}(a), the power density $p_{\rm max}$ increases rapidly
as a function of the hot-side effective convection coefficient, $h_{\rm h}$. At $h_{\rm h} = 10$ W/(m$^2$K), the power density is 1.909 W/m$^2$, which is a fairly good result for still air without assisting structure. It can be asymptotically improved to 2.2 W/m$^2$ {{(with an improvement factor of 0.15)}} for $h_{\rm h}$ near 100 W/(m$^2$K) by a properly designed heat sink or strong wind \cite{ahtt5e}. Conversely, as can be seen in Fig. \ref{fig_3}(b), the power density $p_{\rm max}$ decreases drastically as a function of the cold-side convection coefficient, $h_{\rm c}$. At $h_{\rm c} = 7$ W/(m$^2$K), the generated power density is only 0.1938 W/m$^2$, and asymptotically approaches 2.2 W/m$^2$ for $h_{\rm c}$ near 0.001 W/(m$^2$K). {{The huge improvement factor of 10.35}} is a manifestation of the necessity of a vacuum enclosure \cite{Chen2016} (or special locations as deserts \cite{10.1115/1.3450256}) for achieving the desired power density. 

    \par As already discussed in section \ref{secexper}, the thermoelectric figure of merit ZT has a significant influence on electrical power generation \cite{Snyder2008, C7EE02007D}. Higher ZT amounts to reduced electrical resistance and/or increased thermal resistance of the TEG, the first enhances the output current and the latter reduces the parasitic heating of the cold side. As demonstrated in Fig. \ref{fig_3}(c),  the power density $p_{\rm max}$ increases rapidly as a function of ZT value, {with an improvement factor of 4.542 as ZT increases from 0.71 to 100.} Power densities in the range of Watts/m$^2$ can be obtained with existing TEG \cite{TEG_record} as discussed in the previous section, while new materials of even higher ZT \cite{Shimizu2019_1} - which operate at a temperature range near ambient temperature - which matches with our application, are undergoing development.

    \par In Fig. \ref{fig_3}(d), the power density $p_{\rm max}$ as a function of the area ratio of the TEG to the radiative cooler $A_{\rm TE}/A_{\rm c}$ for three types of emitters is presented. For all of the three, the output power density $p_{\rm max}$ follows a similar trend as in Fig. \ref{fig_4}(b). At the area ratio 0.0286 employed in the experimental set up of \cite{RAMAN20192679}, we predict the optimal emitter power density of 1.5529 W/m$^2$, while the multilayer emitter achieves 1.1169 W/m$^2$ and the blackbody emitter 0.8676 W/m$^2$. For the multi-layer emitter, the maximum output power density $p_{\rm max}$ is  1.2144 W/m$^2$ when $A_{\rm TE}/A_{\rm c}$ is 0.0156, and the blackbody emitter generates the maximal output power density $p_{\rm max}$ of 0.87 W/m$^2$ when $A_{\rm TE}/A_{\rm c}$ is 0.0252. {The improvement factor 0.416 for the optimal emitter is highest among the three emitters, followed by the multilayer emitter 0.087 and the blackbody emitter 0.003. }The very low TEG area compared to the system footprint deserves additional consideration. Naively, from Eq. (\ref{Power_max}), the power density scales linearly with the TEG area (or the number of thermocouples, assuming the area of the thermocouple is fixed), for a given temperature gradient. However, the temperature gradient itself is reduced by increasing the TEG area, since more heat is pumped to the cold side and our cooling resources are limited. The voltage of the TEG, which is proportional to $N(T_{\rm h}-T_{\rm c})$, increases but only as a sublinear function of $N$, while the current which is proportional to $(T_{\rm h}-T_{\rm c})$ decreases monotonically with $N$. The resulting optimal point is thus obtained for a relatively small $A_{\rm TE}/A_{\rm c}$ ratio. 
    \begin{figure}[ht]
\centering\includegraphics[scale = 1.5]{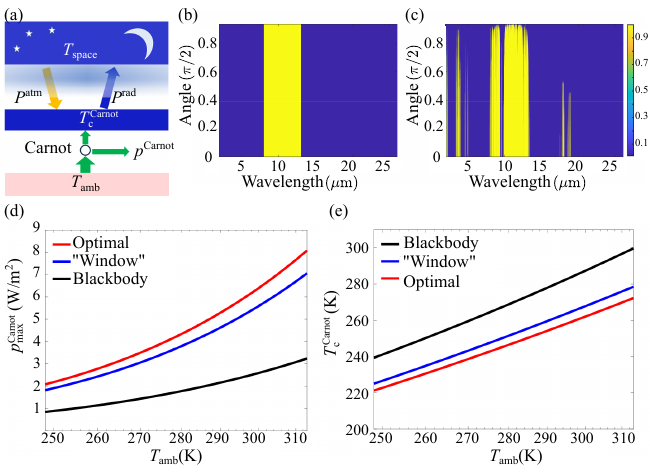}
\caption{Work extracted by a Carnot engine in place of the TEG. (a) Schematics of the Carnot engine (represented by a circular disk) setup. (b) ``Window'' emissivity spectrum. (c) Optimized emitter for maximum generation of work by a Carnot engine operating between the nighttime radiative cooler and the ambient temperature according to Eq. (\ref{optim_emiss}). (d) Thermoelectric power generation performance for blackbody, window and optimal emitters as a function of ambient temperature. (e) Temperature of the cold side for the system in (d).}
\label{fig_5}
\end{figure}
\section{Thermodynamic limit of power generation at night}\label{sec5}
    \par It is important to know the maximal achievable power density from the ambient air during nighttime. Here we again assume a roof top area of 1 m$^2$ with $A_{\rm c} = A_{\rm h} = 1$ m$^2$. In this section, our theoretical analysis is therefore based on replacing
the aforementioned TEG model with a Carnot engine that works between the heat source and sink. Here, we consider a practical atmosphere whose emissivity spectrum is the same as used in the TEG study of previous sections and the radiative cooler emitter is with an emissivity spectrum $\epsilon(\lambda, \theta)$ given by Eq. (\ref{optim_emiss}). Therefore, our study is a departure from the previous papers on the limit for outgoing thermal radiation with idealized atmosphere \cite{Byrnes3927} or without atmosphere \cite{BuddhirajuE3609,Li2020}. In steady state and assuming the net heat flux of the cool side out of the system is only radiative (as is effectively the case in our vacuum enclosed design), the maximum work density extracted by the Carnot engine from the setup of Fig. \ref{fig_5}(a) is, 
\begin{align}
p_{\rm max}^{\rm Carnot} &=\max\limits_{T_{\rm c}^{\rm Carnot}}\left[\left(\frac{T_{\rm amb}}{T_{\rm c}^{\rm Carnot}} - 1\right) \Delta p_{\rm r}(T_{\rm c})\right],
    \label{Carnot_max_balance}\end{align}
where $\Delta p_{\rm r}(T_{\rm c})$ are defined as in Eq. (\ref{lhs_rhs_eq}). For efficient heat flow to the hot side ($h_{\rm h}=10^2$ ${\rm W/(m^2K)}$ as in our optimal design), $T_{\rm h}$ of the Carnot engine can be safely approximated by $T_{\rm amb}$. The value of $T_{\rm c}^{\rm Carnot}$ is optimized in order to maximize Eq. (\ref{Carnot_max_balance}). The corresponding optimal emissivity spectrum at 300 K ambient temperature is shown in Fig. \ref{fig_5}(c). It has similar but not equal spectro-angular selectivity to the TEG spectrum of Fig. \ref{fig_1}(a) as the temperature of the emitter here is lower compared to realistic devices such as TEG.
   \par We compare the thermodynamic limit of power extraction between the optimal emitter and that of the blackbody emitter adopted in \cite{RAMAN20192679} and the window emitter proposed in \cite{Byrnes3927}, as shown in Fig. \ref{fig_5}(b). For the window emitter, we set unity emissivity within the wavelength range from 8 to 13 $\mu$m and zero otherwise as studied in \cite{Byrnes3927}. In Fig. \ref{fig_5}(d), we evaluate the power generation limit for a range of ambient temperatures from 245 K to 315 K. In all three cases, the generated power density limit increases as a function of ambient temperature. Using the optimal emitter results in the highest power density limit. To explain the different performance of these three emitters, we note that the optimal emitter achieves the lowest cooler temperature $T_{\rm c}$ as shown in Fig. \ref{fig_5}(e). At 300 K ambient temperature, the Carnot engine power density generation $p_{\rm max}^{\rm Carnot}$ is $6.4$ W/ m$^2$ with $T_{\rm c}=262.12$ K.

\section{Concluding remarks}\label{sec_6}
We introduced optimal achievable design of nighttime thermoelectric power generation, showing that power density in the range of Watts/m$^2$ is achievable with current technologies. The improved power generation performance is enabled by a spectro-angular-selective emitter. We explored the optimized power generation system with optimal emitters at different thermal convection conditions and TEG parameters. {Starting from the current experimental setup, reducing the cold-side effective convection coefficient has the largest improvement factor, followed by increasing ZT factor and then hot-side convection coefficient, with the weakest influence from changing the area ratio of the TEG to radiative cooler}. We showed that we can achieve performance close to that of the thermodynamic limit set by the Carnot heat engine. This result is significantly higher than the previous reported results and points to the potential applicability of harvesting electrical power at night.

\section*{Acknowledgments}
L.F. acknowledges helpful discussions with Dr. Bo Zhao, Dr. Linxiao Zhu, Dr. Yu(Jerry) Shi, Dr. Siddharth Buddhiraju, Casey Wojcik and Guillermo Angeris. 
\section*{Disclosures}The authors declare that there are no conflicts of interest related to this article. \section*{Funding}
 This work is supported by the U.S. Department of Energy under Grant No. DE-FG02-07ER46426.


\bibliography{OSA-template}

\end{document}